\documentclass[10pt]{article}
\usepackage{amssymb,graphicx}

\newcommand{\eqref}[1]{(\ref{#1})}

\newcommand{\bfn}{}

\newcommand{\be}{\begin{equation}}
\newcommand{\ee}{\end{equation}}

\newcommand{\bea}{\begin{eqnarray}}
\newcommand{\eea}{\end{eqnarray}}

\def\k{{\bf k}}
\def\x{{\bf x}}

\def\lag{\langle}
\def\rag{\rangle}

\def\removetext#1{{}}

\begin{document}

\title{Goldstone excitations from spinodal instability}
\author{
Sz. Bors\'anyi$^{ }$\footnote{e-mail: mazsx@cleopatra.elte.hu}, \ A.
Patk{\'o}s$^{ }$\footnote{e-mail: patkos@ludens.elte.hu} \ and D. Sexty$^{
}$\footnote{e-mail: denes@achilles.elte.hu}\\
 Department of Atomic Physics, E{\"o}tv{\"o}s University,\\
 H-1117 Budapest, Hungary}
\maketitle
\begin{abstract}
The squared mass of a complex scalar field is turned dynamically into 
negative by its $O(2)$-invariant coupling to a real field slowly
rolling down in a quadratic 
potential. The emergence of gapless excitations is studied in real time
simulations  after spinodal instability occurs.  
Careful tests demonstrate that the Goldstone modes appear almost 
instantly after the symmetry breaking is over, much before thermal
equilibrium is established.  
\end{abstract}
\section{Introduction}

This investigation is the first stage of a complex study with the aim to
explore in detail the mechanism of the energy transfer to the Higgs field at 
the end of the inflationary period of 
the Universe as it is anticipated in scenarios of hybrid inflation 
\cite{linde94,garcia98}. 
  
The $O(2)$-symmetric model used in studies of hybrid inflation has
the following Lagrangian density for the real scalar (``inflaton'') field 
$\psi$ and the complex (``Higgs'') field $\Phi=\Phi_1+i\Phi_2$:
\be
L={1\over 2}(\partial_\nu\psi (x))^2-{1\over 2}m_\psi^2\psi^2+
{1\over 2}|\partial_\nu\Phi (x)|^2-{1\over 2}m_\Phi^2
|\Phi |^2-{\lambda\over 24}|\Phi |^4-{1\over 2}g^2\psi^2|\Phi |^2,
\label{lagrange}
\ee
where $m_\Phi^2$ has the wrong sign. 

At the beginning the energy density is carried partly by the inflaton field
due to its $\sim {\cal O}(M_P)$ initial amplitude and partly by the potential
energy of the Higgs-field which starts at the symmetric maximum.  
It is transformed very efficiently into kinetic and gradient energy
densities when the squared effective mass of the Higgs field becomes negative
and the modes of the complex field with 
low spatial frequency $|\k |< |M_{eff}|$ start to increase exponentially due 
to the spinodal instability \cite{felder01}. {\bfn The Higgs field eventually
arrives to a symmetry breaking ground state, on its top with massive and 
massless thermal excitations}. 

The present study was carried out in Minkowski metrics and with classical 
fields starting from an initial state corresponding to the above situation.
Therefore at this stage the results have more relevance to
non-equilibrium  phase transitions in relativistic field theoretical models 
than to cosmology. The results will serve as reference for future
simulations to be performed in an expanding FRW-geometry. Our study was 
inspired by recent papers of Felder {\it et al.} describing also the
evolution of a complex Higgs field after a sudden change of sign of
its squared mass \cite{felder01,kofman01}.

 Very recently a detailed
study concentrating on the statistics of defect formation was
published \cite{copeland02}, where the change of sign in the mass of
the Higgs field is induced by the free motion of a homogenous
real scalar field. {\bfn
 The present work represents a complementary approach 
since we study rather the elementary excitations of the system than extended 
objects.}

Our aim is to investigate systematically the way the different components of 
the complex Higgs field are excited. Special attention
is paid to the real time appearance of the Goldstone mode after the period of
spinodal instabilities. Analogous question was analyzed recently by
Boyanovsky {\it et al.} \cite{boyan99} and by Baacke and Heitmann
\cite{baacke00} in the infinite component (large $N$) limit of the quantum
dynamics. They solved the coupled set of the renormalised equations of motion 
of the order parameter and of the to-be-Goldstone modes following a 
squared mass quench. The effective mass square governing
the mode equations of the Goldstone fluctuations was analyzed which
depends non-linearly on the actual value of the order parameter and
reflects also the global back-reaction of the Goldstone-modes.
{\bfn It was found that the squared mass of the ''pion'' modes starts 
to oscillate around the asymptotic zero value rather early, but
the oscillations are damped only for asymptotically large 
times as $t^{-1}$. 
In a recent paper \cite{baacke02} renormalised non-equilibrium gap 
equations were solved for the effective mass-squares of the longitudinal and 
transversal modes propagating on a time dependent background. These masses were
in earlier papers \cite{nemoto00,verschelde01} with a self-consistent 
parametrisation of the corresponding propagators. 
It turned out that the time-dependent variational mass-squares do not obey 
the Goldstone theorem away from equilibrium.
It will be interesting to compare these semi-analytical
approximate investigations with the direct mass measurements performed in 
numerical simulations of $O(N)$ symmetric systems.} 

The classical (cut-off) field theory provides a useful point of reference, 
since gapless excitations are present in its broken symmetry
phase near equilibrium. Our numerical results for $N=2$
hint at an essentially different picture on the real time genesis of
the Goldstone modes.

The presentation of our results is organised in the following way.
Section 2 summarizes the set of parameters used in and the
algorithmical details applied to the numerical solution. 
Our detailed discussion is divided into two parts. 
In section 3 the methods
for finding the independently moving degrees of freedom are
presented. There we give direct evidence for the early presence of 
massless excitations. In section 4 the process of thermalisation is 
described, pointing out the slow damping of angular O(2) oscillations,
supporting the presence of Goldstone modes. Section 5 contains our
conclusions.

\section{Lagrangian parameters, initial conditions and discretisation}

The parameters we choose for the present investigation imitate
a situation which would be characteristic for a GUT-like transition. 
In lattice units the Higgs
particle has unit squared mass parameter (with the wrong sign), the inflaton
mass square is $10^{-3}$ times smaller. 
The initial amplitude of the inflaton is large:
$\psi_0=15$. {\bfn Variation of its value does not influence
the qualitative features of the time evolution of the system until
its contribution to the potential energy is negligible relative to the
initial energy content of the Higgs field.}

A very weak nonlinearity $\lambda=24\times 10^{-4}$ was fixed in order to 
facilitate the comparison of our results with \cite{kofman01}. In this way 
the dominant part of the initial energy was stored in the Higgs potential. 
When we varied $\psi_0$ in a range, without changing
 the dominant component of the potential energy, no change was experienced 
in the thermalisation scenario after the spinodal instability.
The use of a stronger Higgs self-coupling does not seem to have any qualitative
consequence on the features of the appearance of Goldstone modes.
 
The initial space average of the Higgs field
was fixed to zero. To the two real components of the complex field at
every site a (white noise) amplitude evenly distributed on the
$[-1,1]\times 10^{-4}$ interval was assigned. {\bfn The amplitude
of this random noise can be used to control which phase the system finally 
relaxes to.
With the present choice the final value of $|\Phi|$ is very close to the 
classical minimum of the Higgs potential, the system is very cold. 
One needs 5 order
of magnitude stronger noise to drive the system into the symmetric phase.} 

Runs starting with different initial noise seeds were used to assess the
statistical errors of the mass-values extracted. The restricted number
of runs (6-14 depending on the size of the system), might cause some 
underestimation of the errors.
 
The value of the inflaton-Higgs coupling used in the present 
investigation is $g^2=10^{-2}$. 
It turned out in our numerical studies that with stronger coupling 
(for $g^2\geq 8\times 10^{-2}$) the time necessary for the full development of 
the spinodal instability takes several oscillation periods of the
inflaton, and the growth of unstable modes can happen only
during the passage of the inflaton field through zero. 
This behaviour can be understood with help of the theory of broad
parametric resonance developed in \cite{kofman97} when it is applied to a 
system where also spinodal instability can occur. The main reason for
this phenomenon is the fact that with the increase of $g^2$ the time interval
of the non-adiabatic inflaton evolution becomes shorter. Its
discussion is beyond the scope of the present paper. {\bfn It has been
 checked that the spectral characterisation of the different 
modes as they appear after the spinodal instability are independent of $g^2$.}

We have followed the temporal evolution of the inflaton as well as of
the $O(2)$-field on a spatial lattice with the equations written for 
dimensionless field quantities:

\bea
& 
\ddot \psi(x) - \bigtriangleup \psi(x) + {m_\psi^2 \over m_d^2} \psi(x) 
+ g^2 \Phi^2 \psi =0 \nonumber\\
&
\ddot \Phi_i(x) - \bigtriangleup \Phi_i(x) + { m^2_\Phi \over m_d^2 } 
 \Phi_i(x)
+{ \lambda\over 6 } | \Phi (x) |^2 \Phi_i(x)  
+ g^2 \psi^2 \Phi_i =0 
\eea
with the scaled quantities
\bea
&
{ m^2_\Phi \over m_d^2 }  =-1, \qquad { m^2_\psi \over m^2_d } =0.001,
\nonumber\\
&
{ \psi_d \over m_d } = \psi  ,\quad  { \Phi_d \over m_d } = \Phi ,
\quad  t_d m_d  = t  , \quad a_d m_d = a\equiv 1 
\eea 
(the symbols with index $d$ refer to dimensionfull quantities).

{\bfn The characteristic time scales related to 
the elementary excitations turned out much shorter than the relaxation times,
especially for the long wavelength Goldstone modes. This circumstance
made the choice of the time-step in the standard leapfrog
algorithm used by us a sensitive issue. The stability
of the algorithm was ensured by choosing the time step in the interval
$0.125\geq a_t/a_s\geq 0.0625$.  
The long term stability of the time evolution
 was controlled by monitoring the conservation
of the energy and the $O(2)$ charge of the system. Due to the initial 
conditions the latter was fixed at zero. The charge density for the above
time step was fluctuating around zero with an amplitude $\sim 0.04$, but
in selected runs using hundred times shorter time steps we could reach
an amplitude ${\cal O}(10^{-4})$ with unchanged qualitative features.}

\section{Finding the independent excitations}

The spinodal instability driving the Higgs field from the symmetric phase
to the broken symmetry state was investigated on lattices with increasing
$L^3= 32^3, 64^3, 128^3$ volumes and with periodic
spatial boundary conditions. Due to our initial ``white noise'' 
amplitude distribution the system chooses a well-defined (though
random) direction
in the $O(2)$ internal space contrary to the case studied by \cite{kofman01}
where the space average of the Higgs field stayed at the origin.

\begin{figure}
\begin{center}
\includegraphics[width=9cm]{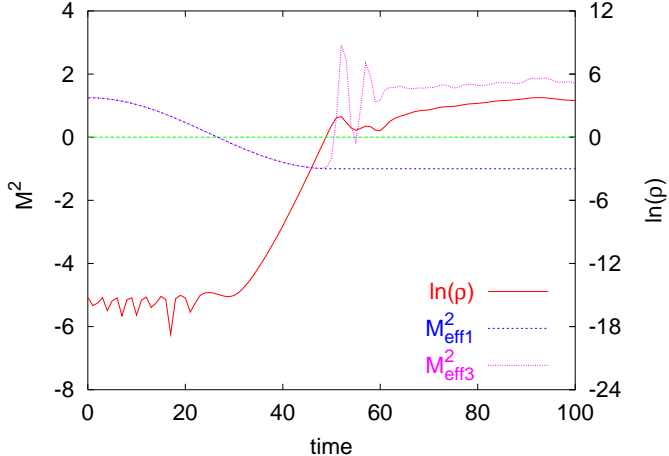}
\end{center}
\caption{Time evolution of $\rho (t)$ (the space average of $|\Phi
  (\x,t)|^2$) and of the various definitions of $M^2_{eff}$ (see
  Eq.(\ref{eff-mass})) }
\label{OP-evol}
\end{figure}

The coincidence of the change of sign of the instant effective squared 
mass $M^2_{eff}$ with the start of the spinodal instability is
investigated in Fig.\ref{OP-evol}. Various alternative definitions of
the effective Higgs mass are used:
\bea
M_{eff,1}^2&=&m_\Phi^2+g^2\lag\psi\rag^2(t),\nonumber\\
M_{eff,2}^2&=&M_{eff,1}^2+g^2\lag (\psi -\lag\psi\rag
)^2\rag,\nonumber\\
M_{eff,3}^2&=&M_{eff,2}^2+{\lambda\over 2}\lag|\Phi|^2\rag.
\label{eff-mass}
\eea
The brackets mean spatial averaging.
The first definition, where one assumes that the fluctuations
of the $\psi$-field are negligible is used in most discussions of hybrid
inflation. It predicts a change of sign when the inflaton reaches the
critical value $\psi_c=|m_\Phi|/g$, and was ensured to be valid in
\cite{copeland02} by setting $m_\psi^2=0$.
In the present case the inflaton  rolls in a harmonic potential
and one can test
if the fluctuations of the $\psi$ and $\Phi$ fields have any effect at
the start of the spinodal instability. In the Figure only $M^2_{eff,1}$ and 
$M^2_{eff,3}$ are displayed, since $M^2_{eff,2}$ is indistinguishable from the 
first one. The deviation of the two versions of the mass appears near
the end of the instability, when $M^2_{eff,1}$ stabilizes at the value $-1$, 
while $M^2_{eff,3}$ 
approaches a positive stationary value after wild oscillations. The 
increase of the order parameter $\rho\equiv \int d^3x|\Phi|/V$ becomes
noticeable with a slight time delay after
the sign-change of the effective Higgs mass. This is due to
the gradual increase of $-M^2_{eff,1}$. The spinodal increase of the
order parameter is exponential, with a slowly varying
slope. Near the saturation of the order parameter the slope is very
close to the unit value expected with our choice of $-m_\Phi^2$.

Next, we turn to the separation of the independently oscillating
degrees of freedom.
At every lattice point the complex field can be represented as $\Phi (\x ,t)=
\rho (\x ,t)\exp (i\varphi (\x ,t))$. The main question of this investigation
was, at what time the radial and angular Higgs variables become
dinamically independent.
This can be monitored by calculating the temporal variation of the  elements of
the ``velocity''-correlation matrix $W_{ij}(t)$, which was constructed the 
following way. 

Let us define the 3-component quantity: $v_i(\x ,t)=\{\dot\rho (\x ,t), 
\rho (\x ,t)\dot\varphi (\x ,t),\dot\psi (\x ,t)\}$ and form the spatial 
fluctuation matrix
\be
W_{ij}={\tilde W_{ij}
\over \tilde W^{1/2}_{ii}\tilde W^{1/2}_{jj}}, \qquad
\tilde W_{ij}(t)={1\over V}\int d^3xv_i(\x ,t)v_j(\x ,t)-
{1\over V}\int d^3xv_i(\x ,t){1\over V}\int d^3xv_j(\x ,t).
\ee
\begin{figure}
\begin{center}
\includegraphics[width=9cm]{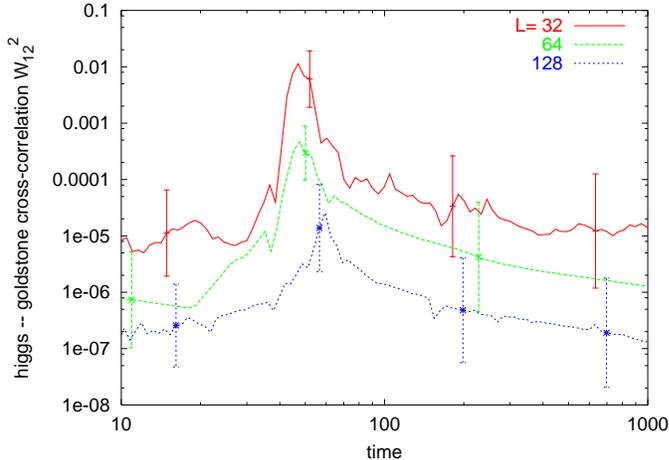}
\end{center}
\caption{Time evolution of the square of the normalised cross-correlation
  coefficient $W_{12}(t)$ on a sequence of different size lattices}
\label{off-corr}
\end{figure}

In Fig.\ref{off-corr} we show the early time variation of the
$W_{12}(t)$ element on three different size lattices.  One observes
that the level of Higgs-Goldstone cross-correlation decreases by about
two orders of magnitude from lattice size $L=32$ to $L=128$. All
non-diagonal correlation coefficients show maxima during the spinodal
instability.  In case of $W_{13}(t)$ it was found that the maximum of
the inflaton-Higgs cross-correlation coefficient is independent of the
size and is ${\cal O}(1)$.  In case of $W_{12}$ and $W_{13} $ also the
maximum decreases with the size of the system.  It is unclear what is
the thermodynamical limit of this maximum reached during the spinodal
instability.

After the saturation of the unstable modes
the average level of all off-diagonal elements of the normalized correlation
matrix stabilizes for $L=128$ at ${\cal O}(10^{-7})$. This level of
decoupling is reached the slowest for $W_{12}$.
Therefore the radial and angular motion of the $O(2)$-field in the
internal space can be considered with very good accuracy as
independent starting from about $t\sim 200$. 
The analysis gives very similar results when applied
directly to the polar coordinates: $x_i=\{\rho, \varphi ,\psi\}$.

This analysis provides the basis for a separate application of various methods
of mass determination to each of the three independent coordinates. 
{\bfn Would the cross-correlations be ${\cal O}(1)$ beyond the instability,
one could not avoid the application of a complicated diagonalisation
process before applying the methods of dispersive mass determination
to be described next.} 

The method developed and tested by us on the example of the
one-component scalar 
field \cite{borsanyi01} has been further improved for the non-trivial task of
the investigation of Goldstone modes. Its present algorithm is summarised
in the following. 

One keeps track of the time evolution of the lower-$k$ spatial Fourier 
transform of all three
$v$-type lattice fields. No binning is applied. Next, one computes the 
frequency-spectra of the time evolution of the power of the single $k$-modes. 
They show several peaks and the strongest one was identified with the 
eigenfrequency of the investigated $\k$-mode
(the others correspond to various forced oscillation modes).
This correspondence yields the dispersion relation $\omega (\k )$. 

\begin{figure}
\begin{center}
\includegraphics[width=9cm]{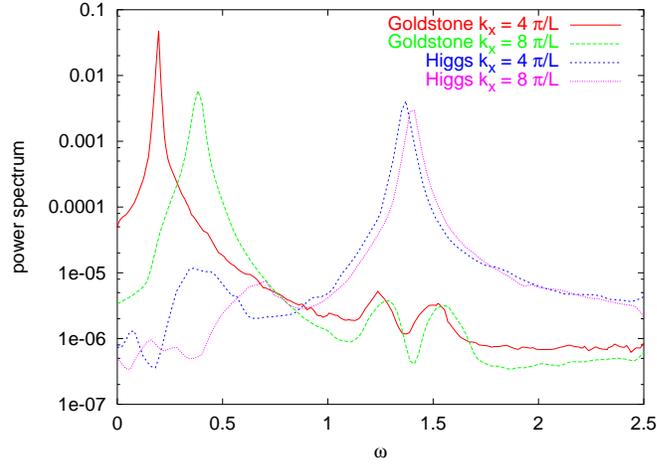}
\end{center}
\caption{Temporal Fourier spectra of the $k_x=4\pi /L, 8\pi /L,
  k_y=k_z=0$ modes derived from long (near equilibrium) time evolution}
\label{omega-spectra}
\end{figure}

In equilibrium systems we verified the correctness of this method by
taking long time intervals. In Fig.\ref{omega-spectra} such
spectra are displayed for the lowest two $\k$ modes for both the Higgs and
the Goldstone variable on the largest $L=128$ lattice.
We observed essentially no volume dependence. The straight line extrapolation
of the dispersion relation from finite $\k$ modes towards zero wave
number is of
very good quality and shows the absence of any gap in case of the angular mode.
The Higgs-mode follows a massive relativistic dispersion relation.

It was checked that the analysis of shorter time
intervals down to $T=32|m_\Phi|^{-1}$ results in masses which are in good
agreement with spectra obtained with much finer $\omega$ resolution.

\begin{figure}
\begin{center}
\includegraphics[width=9cm]{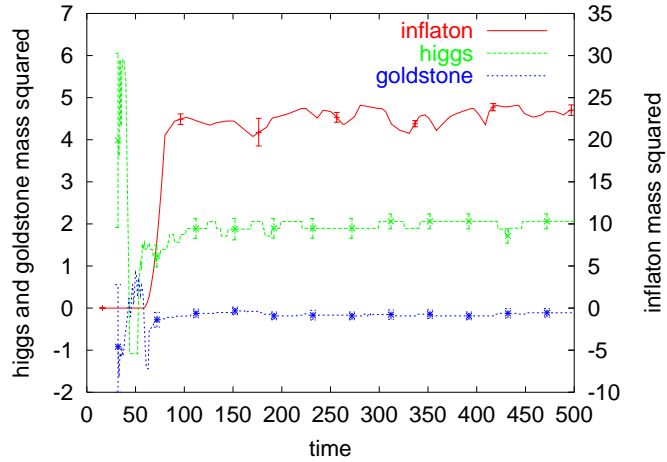}
\end{center}
\caption{The variation of the masses of the Higgs-, Goldstone- and
  inflaton excitations derived from the direct dispersion relation method}
\label{disp-rel}
\end{figure}

Far from equilibrium fields vary fairly fast, any analysis averaged
over a long time interval misses the changes which occur on short
time scales. With the present choice of the parameters a fast Fourier
analysis (FFT) using $T/|m_\Phi|=32$ allows an accurate enough
determination  of the
Fourier coefficients in the interval $\omega\in ( 2\pi/T,12\pi/T)$. 14
independent runs were analysed on an $L=128$ lattice. The masses were
obtained by subtracting the lattice $|\k|^2$ from the corresponding
$\omega^2$ for 9 modes, and fitting the remaining numbers to a
constant. The average standard deviation of the fits performed in
different moments is shown as an error estimate of the mass determination.
 Fig.\ref{disp-rel} displays the time evolution of the independent
inflaton, Higgs and Goldstone masses. 


In an earlier version of this method \cite{borsanyi01} we have assumed that 
all $\k$-modes oscillate with a single eigenfrequency. Then from the ratio 
$|\dot x_i(\k )|^2/|x_i(\k )|^2$ formed with help of $x$-type
variables one can deduce the dominant frequency. In order to minimize
fluctuations the numerator and the denominator were averaged
separately over the modes lying in narrow $|{\k}|$ bins. 
The squared frequency was linearly extrapolated to $|\k|=0$ and the limiting
value identified with the squared mass. 

\begin{figure}
\begin{center}
\includegraphics[width=7.5cm]{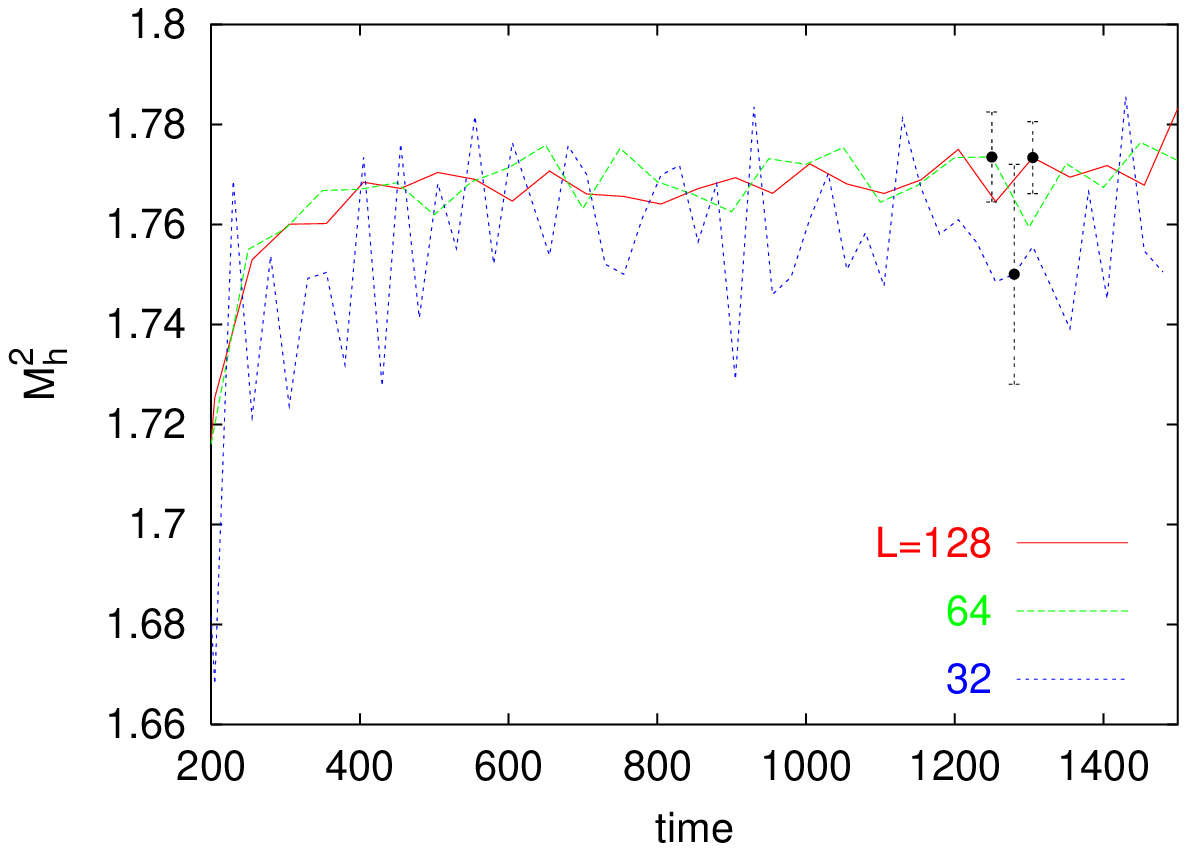}
\includegraphics[width=7.5cm]{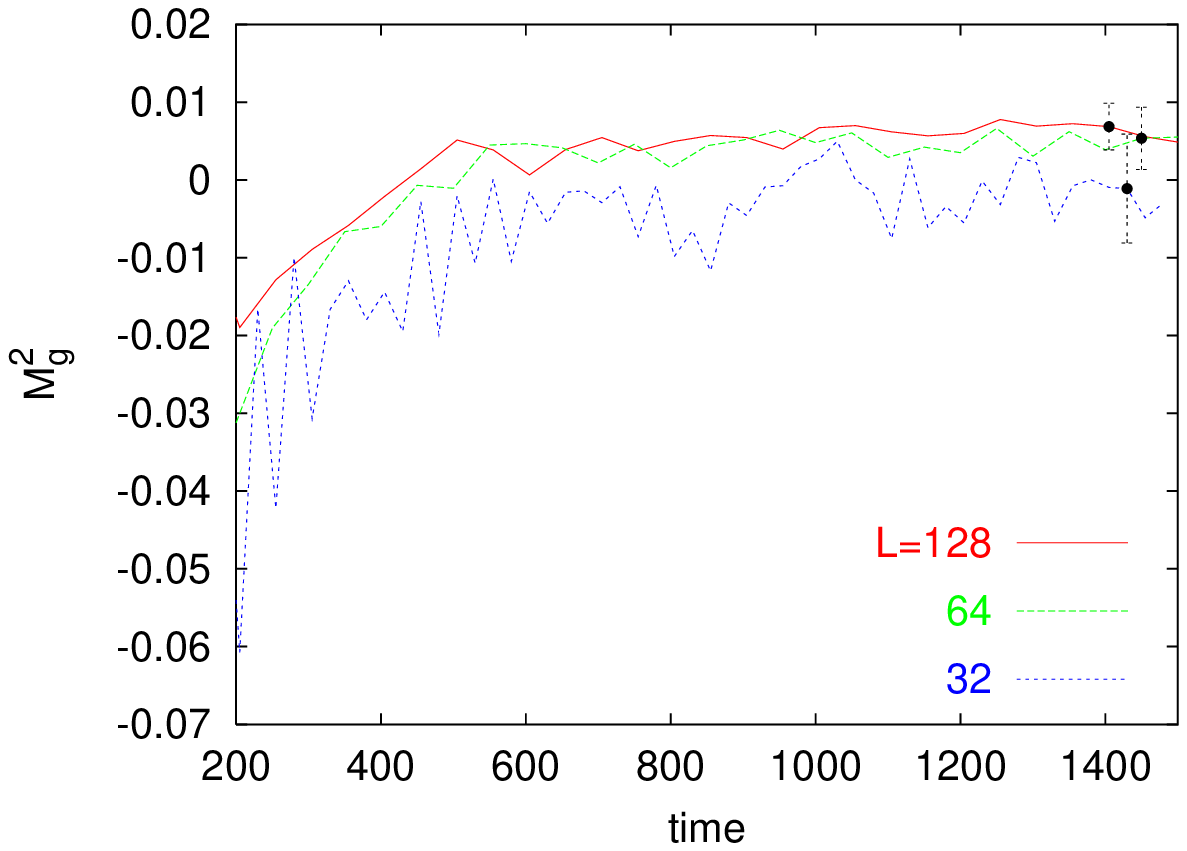}
\end{center}
\caption{Time evolution of the masses of the Higgs (Goldstone) 
excitations derived  from  the ratio of the kinetic to the 
  amplitude power spectra of the corresponding field combinations}
\label{Higgs-one-freq}
\end{figure}

In Fig.\ref{Higgs-one-freq} one can see that the masses
determined with this method provide for both components slightly
higher values than the improved method. Their temporal fluctuations quickly 
diminish with increasing lattice size. The mass estimates shown in
these figures represent averages of 40 independent runs for $L=32$ (9 runs were
analysed for $L=64$, and 6 for $L=128$ with no sizable finite size dependence).
The error bars displayed at the right
hand side of both figures represents the standard deviation of the
squared masses. Within these limitations on
the accuracy the results of the two mass determinations are compatible.
 
For the Goldstone mode 
the best fit was reached by analyzing the phase factor $|\dot X_k|^2/
|X_k|^2,~X=\exp (i\varphi)$. This choice was suggested by the idea of
non-linear $\sigma$-models. We note that only modes with $|\k| \geq
0.5$ were included into the analysis. As we will discuss in the next
section for lower wave numbers the Goldstone modes are highly excited
during and after the instability.
They relax very slowly, and their motion does not seem to follow a
single eigenfrequency. 
From the right hand picture of Fig.\ref{Higgs-one-freq} one
recognizes an apparent (rather small) negative squared mass fit for a
short
time interval directly after the spinodal instability. For later times
the fit is compatible with vanishing mass.

We conclude this section with the statement
that Goldstone modes are present in the excitation spectra of the
model very early. The important question to be discussed in the next
section is the degree of excitation of this mode relative to the
others, that is the part of energy carried by the light modes.

\section{Thermalisation and Goldstone-damping}

\begin{figure}
\begin{center}
\includegraphics[width=9cm]{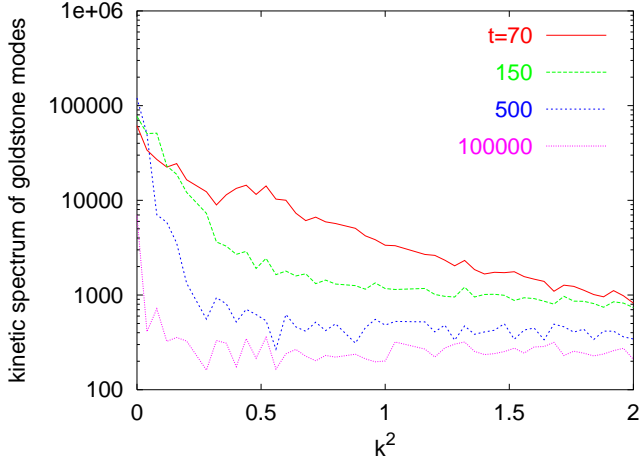}
\end{center}
\caption{The evolution of the Goldstone kinetic power spectrum in time reveals
  the fast approach of high-$k$ modes towards equipartition and the
  slow relaxation in the low-$k$ part of the spectrum}
\label{g-csucs}
\end{figure}

The temporal evolution of the kinetic power spectra reflects very well
the approach to classical thermal equilibrium, characterized by the
equipartition of the energy. In Fig.\ref{g-csucs} the flattening of
the kinetic Goldstone power spectra in the region $k\geq 1$ is clearly
observed. The low $\k$ peak has an almost time independent height in
the interval $t=(100, 1500)$, its decrease is well recognizable only
for $t\sim 10^5$. 

This observation is probably the most appealing result of the present
investigation. Below we attempt its semiquantitative interpretation
 with help of a combination of perturbative and non-perturbative arguments.

In a recent publication (\cite{jakovac00}, Appendix B and C) we have presented
the perturbative evaluation of the Goldstone and Higgs damping rates
in the broken phase of a classical(!) $O(N)$ symmetric field theory.
The analysis was performed for constant value of the order parameter.
The Goldstone and Higgs modes were characterized by constant mass
values and by uncorrelated spectral functions 
\be
\Delta_a(p)= 2\pi\epsilon(p_0)\delta
(p^2-M^2_a)f_a(p_0),\qquad a=H,G.
\label{spectral}
\ee 

The induced source densities for these
equations were found iteratively, assuming weak non-linearities in the
system. The ``self-energy'' terms of the two types of excitations were
established by taking into account the solution with just one iteration.
The following explicit expressions were found:
\bea
\Pi_H(p)&=&(\lambda \Phi)^2\Bigl[R_{HH}(p)+{N-1\over 9}R_{GG}(p)\Bigr],
\qquad \Pi_G(p)=\bigl({\lambda\Phi\over 3}\bigr)^22R_{GH}(p),\nonumber\\
R_{ab}&=&\int{d^4q\over (2\pi)^4}\bigl(\Delta_b(q-p/2)
-\Delta_a(q+p/2)\bigr){1\over 2qp-M_b^2+M_a^2}.
\label{pert-self}
\eea
In equilibrium systems the classical ``number''-distribution is given by
$f_a(q_0)=T/q_0$.
The damping rates are determined by the imaginary parts of the self-energies.

The initial conditions generated by the spinodal instability for the 
later evolution is hard to specify explicitly. In principle one can
attempt the numerical determination of the spectral function of each
of the independent degrees of freedom \cite{aarts01a,aarts01b}. Here
we followed a less systematic path based on
the phenomenological description of the actual system presented in
the preceding section, which confirms the presence of separate Higgs and
Goldstone degrees of freedom with well-defined masses in our $O(2)$
invariant system shortly after the spinodal instability is
over. Therefore we accept for the spectral densities the form of a
free gas for each of them. Then the only remaining task is to find the
nonequilibrium generalisation of the distribution functions $f_a$.

\begin{figure}
\begin{center}
\includegraphics[width=7.5cm]{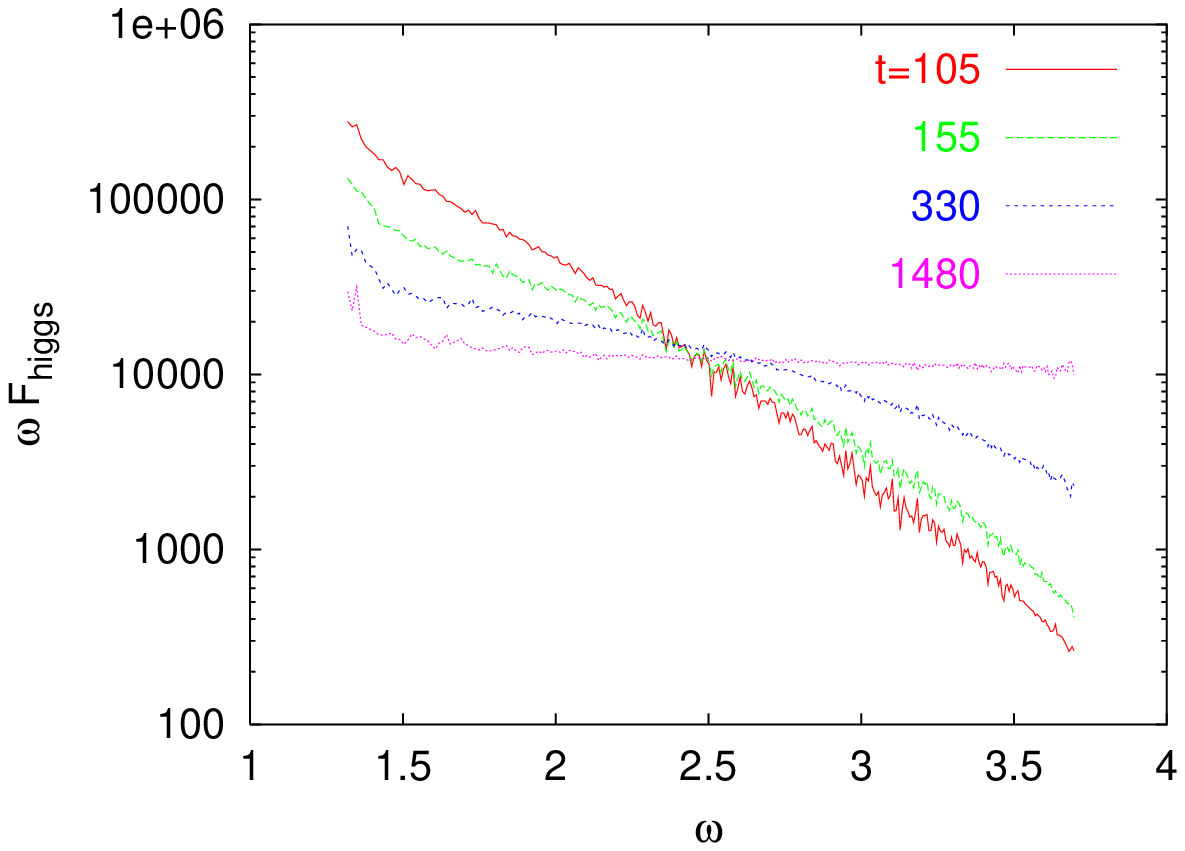}
\includegraphics[width=7.5cm]{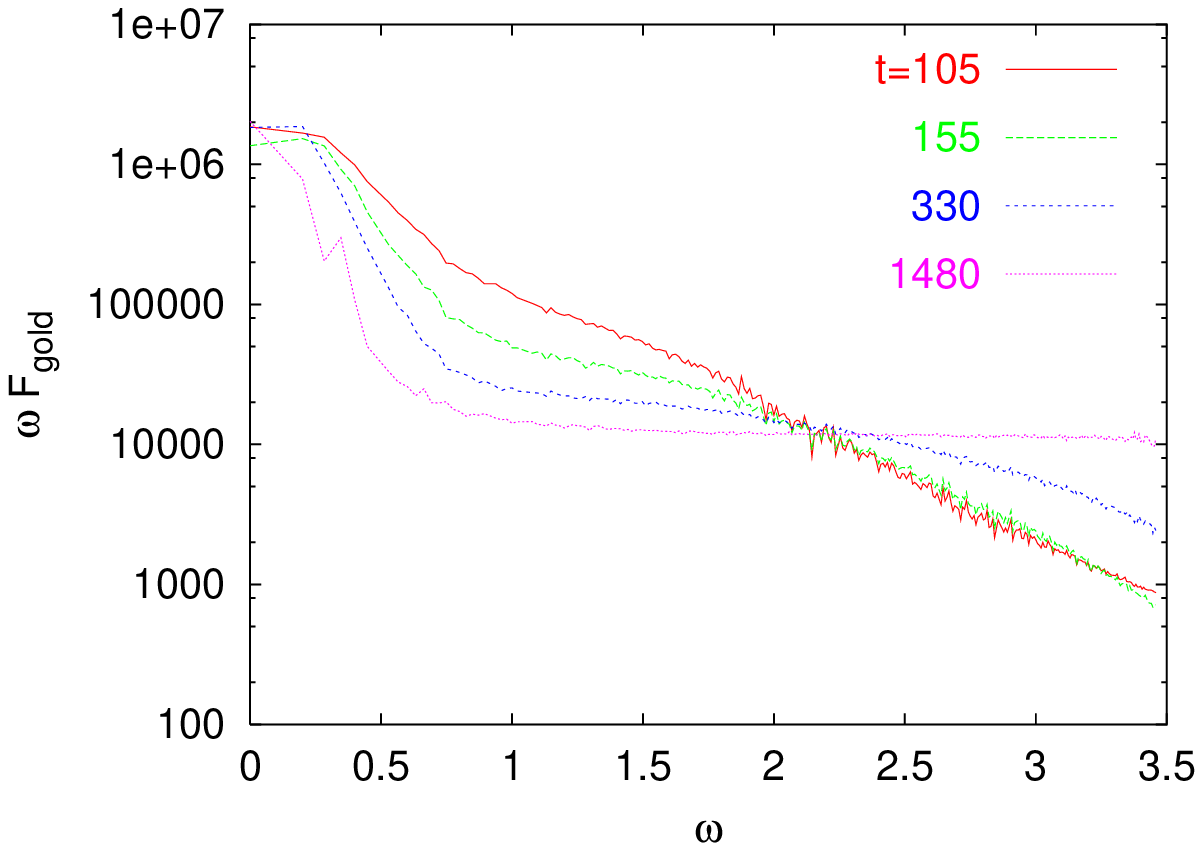}
\end{center}
\caption{The evolution of $\omega\times{\cal F}_a(\omega)$ of the
  Higgs (left) and Goldstone (right) modes shortly after the spinodal
  instability is saturated}
\label{f-distrib}
\end{figure}

 A classical analogue of the average occupation number was extensively used
 for the dynamical coordinate $X_a$ in \cite{kofman01}: 
\be
{\cal F}_a(\omega_k)={1\over \omega_k}(|\dot
X_a(\k)|^2+\omega_a(\k)^2|X_a(k)|^2),
\qquad\omega_a(k)^2=\k^2+M_a^2.
\ee
It was computed here (see Fig.\ref{f-distrib}) for the absolute value
and the phase factor of the complex field. We see
that the ``number density'' of 
the Higgs degree of freedom which initially decreases exponentially
at later times approaches uniformly the $1/\omega_k$ regime.
Also the virial equilibrium of the corresponding kinetic and potential
contributions is fulfilled very early (for $t\sim 500$). 
In case of the Goldstone degree of freedom the convergence to the classical 
``number'' density is not uniform, the excitation of the low-$k$ modes
remains strong, and the approach to the classical regime in the high-$k$ 
region is slower.

The smoothness of ${\cal F}_a$ and the thermalisation tendency observed
in their evolution leads us to the conjecture that the form (\ref{spectral})
of $\Delta_a(p)$ remains
valid in the present case, only the functions $f_a$ are replaced by a 
far-from-equilibrium form.
With this assumption we can use the method of evaluation described in
Appendix C of \cite{jakovac00}. (One should be aware of the fact that
beyond the present one-loop estimation the use of the proposed
non-equilibrium form of $\Delta_a(p)$ would lead to pinch
singularities \cite{altherr94}).  In case of the Goldstone
modes we find the following expression for the on-shell damping rate:
\be
\Gamma_G(k_0=|\k|)=\bigl ({\lambda\Phi\over 3}\bigr )^2{1\over 32\pi|\k|^2}
\int_{M_H^2/4|\k|}^\infty dp_0[{\cal F}_G(p_0)-{\cal F}_H(p_0+k_0)].
\ee

The lower limit is a direct kinematical consequence of the zero mass
nature of the Goldstone mode.
It restricts the contribution to the large
frequency tails of both distributions if the damping rate of the low-$k$ 
modes is to be estimated. Since these modes are filled very late after
the spinodal instability is complete, the damping of these Goldstone modes
is very inefficient.

It is amusing that the far-equilibrium classical evolution under specific 
initial conditions produces a
phenomenon for the Goldstone excitation, which occurs at equilibrium
in the quantum
version of the model due to the exponentially small high frequency tail
of the Bose-Einstein factor \cite{pisarski96,jakovac00,rischke98}.

\begin{figure}
\begin{center}
\includegraphics[width=9cm]{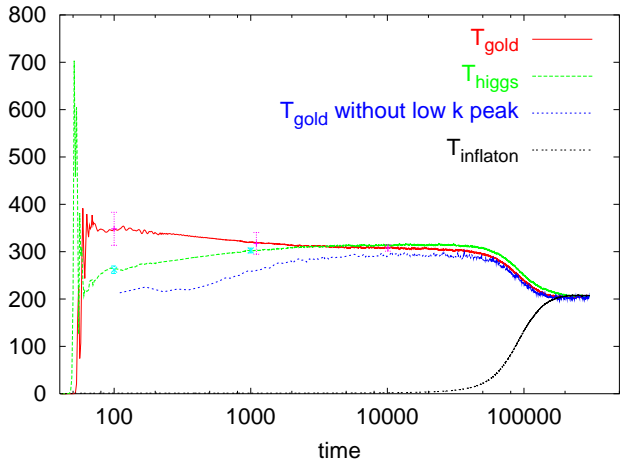}
\end{center}
\caption{The thermal history of the independent degrees of freedom as
  calculated from their average kinetic energies}
\label{therm-hist}
\end{figure}

Due to the rather strong initial
excitation of the low-$k$ Goldstone modes, the average kinetic energy
stored in this degree of freedom is the biggest directly after
spinodal instability. This averaging, however, is somewhat misleading.
By subtracting the energy of the modes participating in the low-$k$
peak, the temperature of the ''equlibrated'' modes appears somewhat
lower than the corresponding Higgs-temperature. With time a metastable
''equilibration'' between these modes becomes almost complete, as one can
see in Fig.\ref{therm-hist}. 

This means that the equilibration in the
$O(2)$ configuration space is much more efficient, than the energy
transfer to the inflaton. Apparently, the inflaton field
is frozen at a much lower temperature than the $O(2)$ fields.

The time spent by the inflaton in frozen state is of the same order of
the magnitude as the relaxation
time of the Goldstone low-$k$ peak. We have checked that with the
increase of $g^2$ the inflaton temperature starts to increase earlier,
but the transition becomes smoother and the equilibrium is reached
later. On the other hand when the complex field is replaced by a real
one  ($g^2$ and
$\lambda$ are kept at the same values) the equilibration time is an
order of magnitude shorter. It is an
interesting question if the final relaxation of the complex field
leading to an equilibration with the inflaton was started by some sort
of instability or it can be characterised by a smooth (exponential
or power law) relaxation.

\section{Conclusions}

In this paper we have shown, that a well-defined Goldstone degree of
freedom appears in an $O(2)$ invariant system, almost instantly after 
the spinodal instability, driven by the evolution of a real field
coupled to it, is
saturated. The thermalisation takes anomalously long due to the very
efficient spinodal excitation of these
modes which showed somewhat unexpectedly extremely slow relaxation
 in the low $k$ region. 

 This feature was understood with help of a leading order iterative
estimation of the Goldstone damping rate relying on the existence of a
far from equilibrium initial spectral function. The fact that a
computation making explicit use of a heavy and 
of a massless field could explain correctly this phenomenon presents
further evidence for the very early presence of the gapless modes
which is the main result of this investigation.

The other surprising effect is the long (metastable) freezing of the
real scalar field $\psi$ coupled to the Higgs mode relatively
strongly. We provided numerical evidence that this phenomenon is
also related
to the presence of the Goldstone modes, but further study is needed
for a complete understanding of the underlying mechanism.

It will be important to repeat the complete study in an expanding space-time
geometry. Recent developments along this line include
the computation of scalar particle production \cite{bellido01} and
the dynamics of coupled real scalar fields \cite{cormier01} in an expanding 
Universe. The efficiency of the late equilibration of the inflaton is
expected to depend critically on the Hubble parameter of the evolution.      
  
During the spinodal instability the low $k$ modes become highly excited,
the classical description is well justifiable for them. 
However, the 
transfer of the energy towards the stable high frequency modes might
be more sensitive to quantum effects. Therefore in a second stage of the
investigation above the upper cut-off $K$ applied to the modes treated 
classically $(K>|m_\Phi| )$, quantum modes will be introduced
 and treated with help of renormalised mode equations \cite{boyan96,baacke97}.

\section*{Acknowledgements}
The authors gratefully acknowledge a valuable suggestion of
D.J. Schwarz and careful reading of the manuscript by A. Jakov\'ac and
Zs. Sz{\'e}p. This investigation was supported by the Hungarian
Research Fund (OTKA).

\end{document}